\documentclass{JHEP3}
\usepackage{amsmath,amssymb,amsfonts,mathrsfs,mathdots}
\usepackage[sort&compress,square,comma,numbers]{natbib}

\newcommand{\D}{\mathrm{D}}
\renewcommand{\d}{\mathrm{d}}

\title{
A worldsheet perspective on string inflation
}
\author{Koenraad Schalm and Ted van der Aalst\\
Institute-Lorentz for Theoretical Physics\\
Niels Bohrweg 2\\
2333 CA Leiden, the Netherlands\\
\email{kschalm@lorentz.leidenuniv.nl, vdaalst@lorentz.leidenuniv.nl}
}

\abstract{We investigate the constraints of slow-roll inflation on the string worldsheet.  A
general gravity-matter set-up is used in which the worldsheet
consists of an
abstract CFT coupled to a 3+1 dimensional non-linear sigma model.
The empirical slow-roll parameters are expressed in terms of the
beta functions of operators in the matter/internal CFT and the
beta function of the dilaton.
The result confirms that inflation is only sensitive to
macroscopic properties of the matter sector, and that in string
theory inflation is a non-perturbative (in $g_s$) phenomenon and
one must go beyond tree-level string theory.}


\begin{document}
\section{Introduction}

The last ten years many attempts have been made to understand
inflation from a more fundamental level within string theory
\cite{Baumann07053837,HenryTye:2006uv,Cline:2006hu,Kallosh:2007ig,Burgess:2007pz,McAllister07102951}. Cosmological observations strongly suggest an era of inflation in the early Universe, and string theory, being a quantum theory of gravity with a unique UV-completion, should be able to describe this. In addition, inflation generically probes energy
scales that are unobtainable in accelerator experiments, and there is a chance that string scale effects may be detectable in future cosmological observations \cite{Martin:2000bv,Kempf:2001fa,Easther:2001fi,Kaloper:2002uj,Greene:2005aj,Jackson:2010cw}.

One of the essential characteristics of inflation is that it
solves the flatness and horizon problem {\em within classical
general relativity} \cite{Guth81,Linde82,Albrecht82}. In string
theory the equations of motion of classical general relativity are
the conditions of conformal invariance of the worldsheet string
theory. Moreover, inflation is a very coarse phenomenon that only
depends on the energy density and pressure in the Universe without
a need to specify any details of the matter content. As such, a
string theoretic description of inflation should only depend on
very generic scaling properties of the conformal field theory on
the worldsheet.


Extending worldsheet descriptions of tachyon condensation
scenarios
\cite{Schmidhuber9404180,Freedman0510126,Hellerman0611317}, we
will attempt to describe inflation with a worldsheet theory that
is a combination of a spacetime and matter-part, which mix via
spacetime dependent couplings $u^a(X)$ for operators
$\mathcal{O}_a$ of an abstract internal CFT. From the viewpoint of
the internal CFT alone such a deformation induces an internal
renormalization group flow. Total conformal invariance of the
combined theory can only be kept if the background fields adjust
themselves in such a way that the running induced by the scaling
behavior of the
operators $\mathcal{O}_a$ of the internal CFT is cancelled. The
renormalization group flow can therefore be seen to define the
possible dependence of $u^a(X)$ on the spacetime coordinates
$X^\mu$, or in other words the beta functions of the full theory
determine the equations of motion for the background fields
$u^a(X)$. These equations can be compared to slow-roll inflation
to find conditions on the internal CFT. We shall indeed find that,
from the worldsheet perspective, the inflationary slow-roll
parameters are completely characterized by the central charge and
the scaling behavior of the couplings of the CFT, in line with our
expectation that inflation is a phenomenon that only depends on
generic properties of the matter sector.

This is not to say that we have solved inflation in string theory.
Describing strings in a time-dependent background is notoriously
difficult. In a large part this is due to our lack of a background
independent description of the theory. At low energies we can
resort to a supergravity description, but inflation fits awkwardly
in the low energy supergravity framework (eta-problem,
Lyth-bound). The worldsheet approach is conceptually different
from supergravity calculations, but it has its own drawbacks when
trying to describe a string in a de Sitter-like background. At
tree-level (in $g_s$), we are only able to describe small
deviations from
{Minkowski} spacetime rather than de Sitter spacetime.
Conversely de Sitter spacetimes cannot be described at string
tree-level, as is well known
\cite{Fischler86,Fischler86II,Das87,Dine85a,Dine85b}. For this reason we may
already anticipate problems to describe inflationary solutions.
Our result is indeed only formal. Substituting the solutions to
the beta functions into the formal expressions, we find a
divergence due to the fact that the dilaton cannot be stabilized
in tree-level string theory (and for a dynamical dilaton inflation
does not occur). This is of course the Fischer-Susskind phenonemon
\cite{Fischler86,Fischler86II}. This, however, is not the main
point. We wish to show that, inflation being a coarse phenomenon,
it only depends on coarse details of the internal CFT. That we do,
while at the same time we recover the known Fischler-Susskind
result that any tree-level string theory model is ruled out as a
theory for inflation.

Our paper is structured as follows: first we describe the
worldsheet set-up suitable for inflation and derive the equations
of motion. We review multi-field slow-roll inflation in section
\ref{sec:multifieldInflation}, so that in section
\ref{sec:wsInflation} we can state our main result. We shortly discuss the possibility to generalize the
results to higher loop order. We conclude discussing the relation between our results
with results known from the literature
\cite{Dine85a,Dine85b}. 


\section{Background dynamics for a generic worldsheet
theory}
\label{sec:eom}

\subsection{Conformal perturbation of a coupled gravity and matter
system}
\label{sec:model}

We wish to describe a realistic model of
inflation in string theory, i.e. there is a $3+1$-dimensional homogeneous and isotropic
cosmological spacetime which inflates. Similar to phenomenological
model building, we are naturally led to consider a worldsheet CFT
consisting of two parts: a non-linear sigma model accounting for
four-dimensional gravity in combination with a matter/internal
theory \cite{Schmidhuber9404180,Freedman0510126}. The non-linear
sigma model is a curved bosonic string in 4 dimensions,
$\mu,\nu\in\{0,1,2,3\}$,
\begin{subequations}
\begin{eqnarray}
S_{NL\sigma M}&=&S_{G(X)}+S_{\Phi(X)},\\
S_{G(X)}&=&\frac{1}{2\pi\alpha'}\int\d^2z\,G^{(S)}_{\mu\nu}\partial X^\mu\bar{\partial} X^\nu,\\
S_{\Phi(X)}&=&\frac{1}{4\pi}\int\d^2z\,\sqrt{g}\Phi(X)R^{(2)}\label{eq:dilaton},
\end{eqnarray}
\end{subequations}
with $G^{(S)}_{\mu\nu}$ the four-dimensional string frame metric
and $g_{\alpha\beta}$ the worldsheet metric. Insisting on
homogeneity and isotropy as in standard cosmology, we set the
Neveu-Schwarz form to zero, $B_{\mu\nu}=0$, but we do consider the
effect of the dilaton. The dilaton is a (light) scalar and is
naturally a part of cosmological dynamics or any time-dependent
scenario, e.g. tachyon condensation
\cite{Freedman0510126}. More importantly, the dilaton is closely
related to the scale factor of the Einstein frame metric and as
such could be driving part of the cosmological expansion.

The internal theory will be some two-dimensional
CFT $S_{CFT_{\mathcal{O}}}$ with
central charge $c_{internal}$ and (primary and descendant) operators
$\mathcal{O}_a$ with scaling dimensions $\Delta_a$.
We purposely leave the theory unspecified. The goal of this paper
is to deduce what type of internal CFT, i.e. which constraints on
the central charge and operator dimensions and couplings, could
give rise to a realistic model for inflation. Since FRW
cosmological dynamics only cares about coarse characteristics of
the matter, viz. pressure and energy, we expect that only coarse
information about the internal CFT should be needed to deduce
cosmological dynamics. Because time-dependent backgrounds must
break supersymmetry, we can incorporate all the fermionic partners
to $X^{\mu}$, and the worldsheet diff$\times$Weyl and
supersymmetry ghosts into the internal CFT.\footnote{One could
keep supersymmetry manifest in principle but it is technically far
more involved: with the worldsheet supersymmetric string one needs
to track the GSO projection carefully whereas the superspace
Green-Schwarz string does not lend itself easily to
non-supersymmetric backgrounds. Essentially all these
technicalities reside in the internal sector and it is not clear
what one would gain by tracking them closely.} The internal CFT
will exhibit characteristic scaling behavior under a deformation
by nonzero couplings $u^a$ to the primary operators,
\begin{subequations}
\begin{eqnarray}
S&=&S_{CFT_{\mathcal{O}}}+S_{\Phi}+S_{u},\\
S_{\Phi}&=&\frac{1}{4\pi}\int\d^2z\,\sqrt{g}\Phi R^{(2)},\\
S_{u}&=&\int\d^2z\,u^a\mathcal{O}_a.
\end{eqnarray}
\end{subequations}
This behavior is intrinsic to the internal theory and fully
captured by the beta functions $\bar{\beta}^a(u)$ of the couplings
$u^a$, whose lowest order (classical) contribution is given by
$(\Delta_a-2)u^a$. We have included the dilaton $\Phi$ here as a
(non $X$-dependent) coupling to the worldsheet curvature $R^{(2)}$
in order to easily incorporate the Weyl anomaly of the theory. At
a renormalization group fixed point of this perturbed CFT,
$\bar{\beta}^\Phi(u)$ will just be proportional to the central
charge of the internal CFT,
\begin{equation}
\bar{\beta}^{\Phi}(u)=\frac{2c_{internal}}{3}+\mathscr{O}(u).
\end{equation}
Due to the conformal perturbations of the internal theory, higher
order effects in $u$ will result in a ``running'' of
$\bar{\beta}^\Phi$
\cite{Zamolodchikov86,Zamolodchikov87}.

To obtain spacetime dynamics driven by the matter sector we couple
the internal theory plus dilaton to the Polyakov non-linear sigma
model into a full worldsheet theory with a cross-coupling
$u^a(X)\mathcal{O}_a$ between the two sectors,
\begin{subequations}\label{eq:fullaction}
\begin{eqnarray}
S_{tot}&=&S_{G(X)}+S_{CFT_{\mathcal{O}}}+S_{\Phi(X)}+S_{u(X)},\\
S_{u(X)}&=&\int\mathrm{\d}^2z\,u^a(X)\mathcal{O}_a.
\end{eqnarray}
\end{subequations}
The couplings $u^a(X)$ to the internal CFT operators
$\mathcal{O}_a$ depend on the spacetime coordinates $X^\mu$.
Since a consistent string theory is described by a conformal
worldsheet theory, the \emph{full} operators $u^a(X)\mathcal{O}_a$
are assumed to be exactly marginal deformations of the theory.
That is the total theory must remain conformally invariant and the
spacetime equations of motion are given by the requirement that
the beta functions of the full theory vanish \cite{Callan85,Hull86,Tseytlin87}.

The beta functions are readily computed using worldsheet
techniques and conformal perturbation theory
\cite{Freedman0510126}. We give a brief summary in appendix
\ref{sec:betacalc}. Here we simply state the result,
\begin{subequations}\label{eq:betaeom}
\begin{eqnarray}
0&=&\frac{1}{\alpha'}\beta^G_{\mu\nu}=R_{\mu\nu}-M_{ab}(u)\nabla_\mu u^a\nabla_\nu u^b+2\nabla_\mu\nabla_\nu\Phi,\\
0&=&\frac{1}{\alpha'}\beta^a=\frac{1}{\alpha'}\bar{\beta}^a(u)-\frac{1}{2}\D\nabla u^a+\nabla^\rho\Phi\nabla_\rho u^a,\\
0&=&\frac{1}{\alpha'}\beta^\Phi=U(u)-\frac{1}{2}\nabla^2\Phi+\left(\nabla\Phi\right)^2,\label{eq:betaeomPhi}
\end{eqnarray}
\end{subequations}
where $M_{ab}(u)$ is the \emph{positive definite} Zamolodchikov metric on the space of
coupling constants\footnote{For later convenience we have rescaled the metric by a
factor of $4\pi^2$ compared to more conventional definitions.} \cite{Zamolodchikov87,Freedman0510126},
\begin{equation}
M_{ab}(u)=4\pi^2\langle\mathcal{O}_a(\epsilon)\mathcal{O}_b(0)\rangle_u;
\end{equation}
we denote its connection by $K^a_{bc}$ and we have defined a
covariant derivative \cite{GrootNibbelink0107272} $\D\nabla u^a$
and scalar function $U(u)$ respectively by
\begin{subequations}
\begin{eqnarray}
\D\nabla u^a&=&\nabla^\rho\nabla_\rho
u^a+K^{a}_{bc}\nabla^\rho u^b\nabla_\rho u^c,\label{eq:covariantderivative}\\
U(u)&=&\frac{2c_{st}}{3\alpha'}+\frac{1}{\alpha'}\bar{\beta}^\Phi(u)
=\frac{2c_{st}}{3\alpha'}+\frac{1}{\alpha'}\bar{\beta}^\Phi(u).
\end{eqnarray}
\end{subequations}
The scalar function $U(u)$ accounts for the different quantum Weyl
anomalous effects. There are contributions from the central
charges of the two components of the theory, $c_{st}=4$ and
$c_{internal}\equiv \frac{3}{2}\bar{\beta}^\Phi(0)$, and in
addition there are higher order effects in $u$, which are
collected in the non-constant parts of $\bar{\beta}^\Phi(u)$.

The actual computation of the beta functions combines two methods
with distinct perturbative expansions: conformal perturbation
theory where $u^a$ and $\delta_a=\Delta_a-2$ are small and
$\bar{\beta}^a(u)=\delta_au^a+\ldots$ is known exactly, and
separately the background field method where $u^a$ can be large
but $\bar{\beta}^a(u)$ and $\nabla u^a$ are required to be small.
By allowing for arbitrary $\bar{\beta}^a(u)$ and
$\bar{\beta}^\Phi(u)$ these methods can be combined in a mixed
$\alpha'$-expansion: it can be made ``exact'' to all orders in
$u^a$, but only to second order in $\nabla u^a$ by capturing all
$u$-dependence in the arbitrary unknown functions $M_{ab}(u)$,
$\bar{\beta}^a(u)$ and $\bar{\beta}^\Phi(u)$. Note that
${\beta}^G_{\mu\nu}(u)$ only depends on $\nabla u^a$  as the two
sectors of the total theory decouple when $u^a$ is
$X$-independent. Limiting ourselves to two derivatives is not an
impediment,  since inflation should be captured by a two
derivative description, especially slow-roll inflation.

\subsection{String dynamics from an action}
The condition for Weyl invariance $\beta^G_{\mu\nu}=\beta^a=\beta^\Phi=0$
determines the  equations of motion for the background fields
$\Phi(X)$, $G_{\mu\nu}(X)$ and $u^a(X)$. A crucial ingredient
for the consistency of this interpretation is the coupling between the
dilaton field $\Phi(X)$ and the other matter fields $u^a(X)$. The potential terms, $\bar{\beta}^a(u)$ and
$\bar{\beta}^\Phi(u)$ in (\ref{eq:betaeom}), are not independent but related
via
\begin{equation}\label{eq:potentialsrelation}
M_{ab}(u)\bar{\beta}^b(u)=\partial_a\bar{\beta}^\Phi(u),
\end{equation}
\emph{to all orders in $u$}. This result may be derived from the fact that the conformal
anomaly $\beta^\Phi$ is a $c$-number rather than an operator by the
Wess-Zumino consistency condition
\cite{Curci87,Tseytlin87,Callan85,Callan89,Freedman0510126}. In
particular $\beta^\Phi$ is $X$-independent and hence
$\nabla_\mu\beta^\Phi$ vanishes. Since
$\beta^G_{\mu\nu}=\beta^a=0$, we can verify
\begin{eqnarray}
0&=&-\frac{4}{\alpha'}\nabla_\nu\beta^\Phi=-\frac{4}{\alpha'}\nabla_\nu\left(\beta^\Phi-\frac{1}{4}\beta^{G\mu}_{\phantom{G\mu}\mu}\right)
=\frac{4}{\alpha'}\left(M_{ab}\bar{\beta}^b-\partial_a\bar{\beta}^\Phi\right)\nabla_\nu
u^a.
\end{eqnarray}
The last step follows from the explicit formulae for the
beta functions (\ref{eq:betaeom}). Recall that the beta functions are derived up to second order in $\nabla u^a$ but are \emph{exact} in powers of zeroth derivatives of $u$ due to
the incorporation of all zeroth derivatives of $u$ in the
potential functions $\bar{\beta}^a(u)$ and $\bar{\beta}^\Phi(u)$.
Whereas our result is only an effective description for the
connection between spacetime and matter sector, the matter sector
itself is described exactly.

As a result of the relation (\ref{eq:potentialsrelation}) between
$\bar{\beta}^\Phi(u)$ and $\bar{\beta}^a(u)$ the equations of motion can be integrated to an action
\begin{equation}\label{eq:SFinflationaryaction}
S_{SF}=\frac{1}{2\kappa_0^2}\int\d^4
x\sqrt{-G}e^{-2\Phi}\left[R+4(\nabla\Phi)^2-M_{ab}\nabla_\mu
u^a\nabla^\mu u^b-4U(u)\right].
\end{equation}
Transforming to the Einstein frame $\tilde{G}^{(E)}_{\mu\nu}=e^{\Phi_0-\Phi}G^{(S)}_{\mu\nu}=e^{-\tilde{\Phi}}G^{(S)}_{\mu\nu}$,
we obtain an action that can be directly compared to standard cosmological models,
\begin{equation}\label{eq:EFinflationaryaction}
S_{EF}=\frac{1}{2\kappa^2}\int\d^4x\sqrt{-\tilde{G}}\left[\tilde{R}-2\tilde{\nabla}_\mu\tilde{\Phi}\tilde{\nabla}^\mu\tilde{\Phi}-M_{ab}\tilde{\nabla}_\mu
u^a\tilde{\nabla}^\mu u^b-4e^{2\tilde{\Phi}}U(u)\right].
\end{equation}
Here $\kappa=\kappa_0e^{\Phi_0}=\sqrt{8\pi G_{N}}$ is the
gravitational coupling. The action (\ref{eq:EFinflationaryaction}) is simply that of a multi-scalar field model coupled to gravity,
\begin{eqnarray}
S_{inflation}&=&\frac{1}{\kappa^2}\int\d^4x\sqrt{-G}\left[\frac{1}{2}R-\frac{1}{2}g_{ij}\partial^\mu\phi^i\partial_\mu\phi^j-V(\phi)\right],\
\label{eq:inflation}
\end{eqnarray}
with the potential
\begin{eqnarray}
V(\phi)&=&2e^{-2\Phi_0}e^{2\Phi}U(u),\label{eq:scalarpotential}
\end{eqnarray}
where we have defined a multi-scalar field  $\phi^i=(\Phi,u^a)^t$ and a metric on the space of fields
$g_{ij}=\left(\begin{smallmatrix}2&0\\0&M_{ab}\end{smallmatrix}\right)$. Since we will be working in the Einstein frame from here on, we have dropped the 
tilde on the spacetime metric $G_{\mu\nu}(X)$. The question we wish to investigate is
whether the potential (\ref{eq:scalarpotential}) is flat enough to
provide realistic slow-roll inflation. Since $V(\phi)$ is proportional to the
beta function $\bar{\beta}^\Phi(u)$ of the internal sector and the
central charge $c$ of the total theory,
demanding slow-roll inflation is equivalent to a set of phenomenological constraints on the internal conformal field theory. Before we turn to this question, we quickly review slow-roll inflation in multi-field models.

\section{Multi-field slow-roll inflation}\label{sec:multifieldInflation}

The rapid acceleration of the universe that characterizes
inflation arises when the system is potential energy dominated.
Current observations favor an adiabatic slow-roll inflationary
model of early universe cosmology, whose phenomenology can be
described by gravity coupled to a single scalar field. The single
field inflationary case was formalized in
\cite{Liddle9408015}. Fundamentally there is no reason to have only one scalar field. Indeed in string theory or supergravities one generically has multiple scalar fields, although its characteristic singature, isocurvature fluctuations, is at most 10\% of the primordial powerspectrum and is at this time not a better fit to the data \cite{Komatsu:2010fb}.  The connection to the powerspectrum for multi-field slow-roll inflation \cite{Linde82,Albrecht82,Kodama84,Starobinsky85} was formalized
in \cite{GrootNibbelink0107272,Langlois:2008qf,Langlois:2008wt}. We shall follow \cite{GrootNibbelink0107272}.

Minimally coupled multi-field inflation is described by the action
(\ref{eq:inflation}), where $V(\phi)$ is the scalar potential and
$g_{ij}$ is the
\emph{positive definite} metric on the space of scalar fields.
For a homogeneous and isotropic FRW universe, 
the independent equations of motion for the generic multi-field action (\ref{eq:inflation}) are\footnote{There is another equation of motion, $\dot{H}=-|\dot{\phi}|^2/2$, from the spatial part of the variation with respect to the metric, but this also follows from (\ref{eq:scalarfieldeqns}).}
\begin{subequations}\label{eq:scalarfieldeqns}
\begin{eqnarray}
H^2&=&\frac{1}{3}\left(\frac{1}{2}g_{ij}\dot{\phi}^i\dot{\phi}^j+V\right),\\
0&=&\D\dot\phi^i+3H\dot{\phi}^i+g^{ij}\partial_j V,
\end{eqnarray}
\end{subequations}
where $H$ is the Hubble parameter, $\Gamma^i_{jk}$ are the connection coefficients for the metric $g_{ij}$ and where we define
\begin{equation}
\D\dot\phi^i=\ddot\phi^i+\Gamma^i_{jk}\dot\phi^j\dot\phi^k,
\end{equation}
similar to (\ref{eq:covariantderivative}).
%
%
The field equations (\ref{eq:scalarfieldeqns}) completely
determine the dynamics of the model. Unfortunately equations
(\ref{eq:scalarfieldeqns}) are difficult to solve exactly and one
generally studies them in a slow-roll approximation. To this end
slow-roll parameters are defined
\cite{GrootNibbelink0107272},\footnote{The definition for $\eta$
differs from the definition $\tilde{\eta}=V''/V$ normally used in
single field slow-roll inflation. Specified to the single field
case, $\tilde{\eta}$ and $\eta$ are related by
$\eta^\parallel=\epsilon-\tilde{\eta}$ to lowest order in the
slow-roll approximation, cf.~(\ref{eq:SRpara}).}
\begin{equation}\label{eq:SRdef}
\epsilon=-\frac{\dot{H}}{H^2},\qquad\eta^i=\frac{\D\dot\phi^i}{H|\dot{\phi}|}.
\end{equation}
The vector $\boldsymbol{\eta}$ can be decomposed in components
parallel $\eta^\parallel$ and perpendicular $\eta^\bot$ to the
field velocity $\boldsymbol{\dot\phi}$. Define
\begin{equation}
\boldsymbol{e}^i_1=\frac{\dot\phi^i}{|\dot\phi|},\qquad\boldsymbol{e}^i_2=\frac{\D\dot\phi^i-\frac{\D\dot\phi\cdot\dot\phi}{|\dot\phi|^2}\dot\phi^i}{\left|\D\dot\phi-\frac{\D\dot\phi\cdot\dot\phi}{|\dot\phi|^2}\dot\phi\right|},
\end{equation}
then
\begin{equation}
\eta^\parallel=\boldsymbol{e}_1\cdot\boldsymbol{\eta}=\frac{\D\dot\phi\cdot\dot\phi}{H|\dot\phi|^2},\qquad
\eta^\bot=\boldsymbol{e}_2\cdot\boldsymbol{\eta}=\frac{\left|\D\dot\phi-\frac{\D\dot\phi\cdot\dot\phi}{|\dot\phi|^2}\dot\phi\right|}{H|\dot\phi|},
\end{equation}
and
\begin{equation}
\eta^i=\eta^\parallel\boldsymbol{e}^i_1+\eta^\bot\boldsymbol{e}^i_2.
\end{equation}

The parameter $\epsilon$ is a direct measure for inflation, since \cite{Liddle9408015}
\begin{equation}
\ddot{a}>0\qquad\Leftrightarrow\qquad\epsilon<1.
\end{equation}
Furthermore $\epsilon$ and $\eta$ together quantify the relative
energy contributions of kinetic and potential energy. One can
reexpress (\ref{eq:scalarfieldeqns}) in terms of the slow-roll
parameters,
\begin{subequations}\label{eq:scalarfieldeqnsinslowrollparameters}
\begin{eqnarray}
H^2&=&\frac{V}{3}(1-\frac{1}{3}\epsilon)^{-1},\\
\dot{\phi}^i+\frac{1}{\sqrt{3V}}g^{ij}\partial_j V&=&-\frac{1}{3}\sqrt{\frac{2}{3}}\frac{\sqrt{\epsilon V}}{1-\frac{1}{3}\epsilon}\left(\eta^i+\frac{\epsilon\frac{\dot{\phi}^i}{|\dot{\phi}|}}{1+\sqrt{1-\frac{1}{3}\epsilon}}\right).
\end{eqnarray}
\end{subequations}
As it is given here equation
(\ref{eq:scalarfieldeqnsinslowrollparameters}) is exact. It shows
precisely which approximation is made by assuming that ``potential
energy strictly dominates over kinetic energy'', which is often the
explanation behind slow-roll inflation. Using
(\ref{eq:scalarfieldeqnsinslowrollparameters}) one could also obtain
results which are any order in slow-roll
\cite{Liddle9408015,GrootNibbelink0107272}. Limiting ourselves to first order in the
approximation, in which
$\epsilon,\sqrt{\epsilon}\eta^\parallel,\sqrt{\epsilon}\eta^\bot\ll
1$, equation (\ref{eq:scalarfieldeqnsinslowrollparameters})
reduces to
\begin{subequations}
\begin{eqnarray}
H^2&=&\frac{1}{3}V,\\
\dot{\phi}^i&=&-\frac{1}{\sqrt{3V}}g^{ij}\partial_j V.
\end{eqnarray}
\end{subequations}
The second equation tells us that slow-roll approximation implies
\emph{gradient flow}. Using these equations we see
that in the slow-roll approximation
\begin{subequations}
\begin{eqnarray}
\dot{H}&=&\frac{1}{2H}\frac{1}{3}\dot{V}=\frac{1}{6\sqrt{\frac{V}{3}}}\partial_iV\dot{\phi}^i=-\frac{\sqrt{3}}{6\sqrt{V}}\frac{1}{\sqrt{3V}}g^{ij}\partial_iV\partial_j V=-\frac{1}{6V}|\nabla V|^2,\\
\D\dot\phi^i&=&\partial_t\left(-\frac{1}{\sqrt{3V}}g^{ij}\partial_j V\right)+\Gamma^i_{jk}\frac{1}{3V}g^{jl}g^{km}\partial_lV\partial_mV=\frac{1}{6}\nabla^i\frac{|\nabla V|^2}{V},
\end{eqnarray}
\end{subequations}
and hence in the slow-roll regime,
\begin{subequations}\label{eq:SRpara}
\begin{eqnarray}
\epsilon&=&-\frac{\dot{H}}{H^2}=\frac{1}{2}\frac{|\nabla V|^2}{V^2},\\
\eta^i&=&\frac{\D\dot\phi^i}{H|\dot{\phi}|}=\frac{1}{2\left|\nabla V\right|}\nabla^i\frac{|\nabla V|^2}{V},\\
\eta^\parallel&=&\frac{\D\dot\phi\cdot\dot\phi}{H|\dot\phi|^2}=\frac{-1}{2|\nabla V|^2}\nabla V\cdot\nabla\frac{|\nabla V|^2}{V}= \epsilon- \frac{\nabla^iV\nabla^jV\nabla_{i}\nabla_jV}{V|\nabla V|^2},\\
\eta^\bot&=&\frac{\left|\D\dot\phi-\frac{\D\dot\phi\cdot\dot\phi}{|\dot\phi|^2}\dot\phi\right|}{H|\dot\phi|}=\frac{1}{2|\nabla V|}\sqrt{\left|\nabla\frac{|\nabla V|^2}{V}\right|^2-\frac{\left(\nabla V\cdot\nabla\frac{|\nabla V|^2}{V}\right)^2}{|\nabla V|^2}} \\
\nonumber
&=&\sqrt{\frac{1}{4|\nabla V|^2} \left|\nabla\frac{|\nabla V|^2}{V}\right|^2-(\eta^{\parallel})^2}.
\end{eqnarray}
\end{subequations}

\section{Inflation from the worldsheet}\label{sec:wsInflation}
\subsection{Slow-roll parameters for tree-level worldsheet string
theory}\label{sec:NScosmo}

We are now in a position to address our question: how do we
describe slow-roll inflation in terms of worldsheet dynamics? That
is, we need to verify that the potential
$V(\Phi,u)=2\left(\frac{\kappa_0}{\kappa}\right)^2e^{2\Phi}U(u)$
is capable of driving a slowly rolling inflaton field. We shall
assume the spacetime part of the worldsheet theory to describe an
accelerating (i.e. inflationary) flat, homogeneous and isotropic
FRW universe, $G^{(E)}=\mathrm{diag}(-1,a^2(t),a^2(t),a^2(t))$,
which is driven by a homogeneous dilaton $\Phi(X)=\Phi(t)$ and
homogeneous internal fields $u(X)=u(t)$. The demand that the
slow-roll parameters are small then provides restrictions on
$V(\Phi,u)$ and hence, as conjectured, on the coarse
characteristics of the internal CFT, $c$, $\bar{\beta}^\Phi(u)$
and $\bar{\beta}^a(u)$. Direct calculation of (\ref{eq:SRpara})
for $V(\phi)=2\left(\frac{\kappa_0}{\kappa}\right)^2e^{2\Phi}U(u)$
reveals
\begin{subequations}
\begin{eqnarray}
\epsilon&=&1+\frac{1}{2}\gamma^2,\label{eq:SRparaepsexplicit}\\
\eta^\parallel&=&-\epsilon-\frac{D}{2+\gamma^2},\\
\eta^\bot&=&\sqrt{\frac{1}{4}(2+\gamma^2)^2+D+\frac{\frac{\gamma^b\gamma^c\nabla_a\bar{\beta}_b\nabla^a\bar{\beta}_c}{\alpha'^2U^2}-2\gamma^2D-2\gamma^6+\gamma^4}{2+\gamma^2}-(\eta^\parallel)^2},
\end{eqnarray}
\end{subequations}
where we have defined the combinations,
\begin{subequations}
\begin{eqnarray}
\gamma_a(u)&=&\frac{M_{ab}\bar{\beta}^b}{\alpha'U}=\partial_a\ln U=\partial_a\ln
\left[\frac{2c_{st}}{3\alpha'}+\frac{1}{\alpha'}\bar{\beta}^\Phi\right],\\ D&=&\frac{\gamma^a\gamma^b\nabla_a\nabla_b U}{U}-\gamma^4.
\end{eqnarray}
\end{subequations}
From (\ref{eq:SRparaepsexplicit}) we immediately see that
$V(\Phi,u)=2\left(\frac{\kappa_0}{\kappa}\right)^2e^{2\Phi}U(u)$
is incapable of driving inflation: $\epsilon$ is always larger
than unity. Regardless of the specific form of
$\gamma_a=\partial_a\ln
\left[\frac{2c_{st}}{3\alpha'}+\frac{1}{\alpha'}\bar{\beta}^\Phi\right]$, the positive
definiteness of the Zamolodchikov metric $M_{ab}$ ensures that
$\gamma^2>0$.

Tracing back we see that the coefficient $1$ in $\epsilon$,
characteristic of an exponential potential, is due to the dynamics
of the dilaton. One could wonder whether taking $\Phi$ constant,
i.e. excluding it from the cosmological dynamics, would modify the
model into one which does allow for inflation. Because the field
space metric $g_{ij}$ is block diagonal, equation
(\ref{eq:scalarfieldeqns}) implies that for a constant $\Phi$,
$\Phi$ must be stabilized at $\partial_\Phi
V=4\left(\frac{\kappa_0}{\kappa}\right)^2e^{2\Phi}U=0$. However,
excluding $\Phi=-\infty$,
 the relation (\ref{eq:potentialsrelation}) precludes a constant dilaton, as $U$ is not allowed to vanish. In our set-up fields $u^a(X)$ that undergo a time evolution in four-dimensional spacetime are described by a renormalization
group flow of the couplings, i.e. $\bar\beta^a\neq 0$. Equation
(\ref{eq:potentialsrelation}) then implies that $U$
cannot vanish, which forces the dilaton to be non-constant by the
requirement (\ref{eq:betaeomPhi}) of a vanishing $\beta^\Phi$.
Turning the argument around, suppose one magically stabilizes the
dilaton at tree-level. Then
$\epsilon=U^{-2}\bar{\beta}^a\bar{\beta}_a$ but $U\sim
\partial_\Phi V$ which must vanish by the assumption that the
dilaton is stabilized.

Within tree-level worldsheet string
theory, the dilaton is therefore always part of the cosmological dynamics and its tree-level exponential potential rules out an inflationary universe. 

\subsection{Inflation from the Ramond sector, string loop corrections and inflation from open strings}

Clearly to describe inflation in string theory we must have a more
complicated potential for the dilaton. One guess could be to
supersymmetrize the worldsheet and include RR fields. Technically
this is a far from trivial task, as it is not yet known how to
compute beta functions for RR vertex operators. However at the end
of the day, even including fermionic dynamics, the resulting
worldsheet theory must be of the form (\ref{eq:fullaction}). On
the worldsheet the dilaton/vertex-operator interactions are such
that they always lead to an action $S=\int e^{-2\Phi}\mathcal{L}$
in the string frame \cite{Polchinski98}. Thus one always deduces
equation (\ref{eq:SFinflationaryaction}) and the remainder of the
analysis is the same.


Thus one is naturally led to consider string loop
corrections or non-perturbative effects, i.e. open strings. From the worldsheet point of view these two
additions roughly boil down to the same thing. Both are
obtained by including more general worldsheet topologies than just
the spherical worldsheet of tree-level string theory.
The corrections from including closed string loops could convert
$\epsilon$ into a more sensible expression. We can expect this
based on the well-known dilaton tadpoles of Fischler-Susskind
\cite{Fischler86,Fischler86II}. Our results are an
extension of the Fischler-Susskind result that to obtain a
worldsheet description of strings in a de Sitter space, there must
be a one loop (in $g_s$) contribution to the dilaton to have
vanishing beta functions, i.e. to satisfy the equations of motion.
Slow-roll inflation is in essence an adiabatic continuation of de
Sitter space to a slowly varying vacuum energy.

It is interesting to see what happens if we suppose that the
higher loop contributions allow us to consistently stabilize the
dilaton at weak coupling independent of the value of $u^a$. Then
one finds the slow-roll parameters
\begin{subequations}
\begin{eqnarray}
  \label{eq:1}
 \epsilon &=& \frac{\bar{\beta}^a\bar{\beta}_a}{2(\bar{\beta}^{\Phi}+ \frac{2c_{st}}{3})^2}, \\
  \eta^{\parallel} &=&\epsilon - \frac{\bar{\beta}^a\bar{\beta}^b\nabla_a\bar{\beta}_b}{(\bar{\beta}^{\Phi}+\frac{2c_{st}}{3})\bar{\beta}^c\bar{\beta}_c}, \\
\eta^{\perp} &=& \sqrt{\frac{1}{4\bar{\beta}^c\bar{\beta}_c}\left|\nabla_a\frac{\bar{\beta}^b\bar{\beta}_b}{(\bar{\beta}^{\Phi}+\frac{2c_{st}}{3})}\right|^2 - (\eta^{\parallel})^2 }.\end{eqnarray}
\end{subequations}
The dilaton stabilization needs to be such that
$\alpha'U=\bar{\beta}^{\Phi}+ \frac{2c_{st}}{3}$ is no longer
proportional to $\partial_{\Phi}V$ and hence the above expressions
make sense. 
Of course dilaton stabilization at weak coupling has its own
problems \cite{Dine85a,Dine85b}.

The inclusion of open strings, in addition to the closed strings
considered here, may yield more promising results for describing
worldsheet theories on inflationary backgrounds. In the
supergravity literature the usefulness/necessity of open string
corrections has already been recognized
\cite{Douglas0610102,Kachru0301240,Baumann07053837,HenryTye:2006uv,Cline:2006hu,Kallosh:2007ig,Burgess:2007pz,McAllister07102951}.\footnote{In
supergravity language open strings add D-terms in addition to
F-term potentials. The closed string worldsheet only captures a
dilaton type  F-term inflation.} Open strings have been
extensively investigated from a low energy effective field theory
point of view, e.g. DBI inflation, and all known viable
supergravity inflationary models have an open string component.

\section{Conclusion}
Inflation does not care about anything but
very coarse features of the matter sector, only its pressure and energy. This suggests that in string theory
inflation is determined by coarse features of the internal conformal field theory on the worldsheet. Qualitatively this is what we find.
At the same time our result shows that 
it is not possible to have an inflationary cosmology described by a tree-level string
worldsheet. The exponential potential for the dilaton ensures that $\epsilon$ is strictly larger than unity, completely independent of the internal CFT. 
At first sight this conclusion may be puzzling, as inflation is a
classical phenomenon and one therefore may expect tree-level
string theory to be sufficient for a consistent description.
Nevertheless the result simply recovers that de Sitter backgrounds
are known to only arise at one-loop level in worldsheet string
theory through the Fischler-Susskind mechanism
\cite{Fischler86,Fischler86II,Das87}. For inflation to occur the dilaton must be stabilized through such higher loop effects. If this stabilization happens at weak coupling, then inflation is possible with slow-roll parameters that only depend on the beta functions of the internal CFT.

%

In a way Fischler-Susskind and the result here are special cases
of Dine-Seiberg runaway \cite{Dine85a,Dine85b}: within string
theory one cannot probe a nearby vacuum from the original vacuum
because in string perturbation theory as currently understood all
higher order corrections are larger than the first order ---
string theory is either free or strongly coupled. Whereas the
result in
\cite{Dine85a,Dine85b} is obtained by general reasoning, Fischler-Susskind specifically attempt to describe a de Sitter cosmology from a Minkowski worldsheet, and we attempt to obtain inflation.
We can be even more explicit: in our tree-level analysis the
time-dependent process of inflation requires on the one hand a
non-constant dilaton to satisfy the equations of motion, while on
the other hand only a constant dilaton makes sense
observationally. In the tree-level limit we therefore have found a
clear inconsistency of the approach. A strong coupling analysis is
necessary to realize inflation within string theory. The reader
should be aware that we have
\emph{not} ruled out a non-constant dilaton scenario at all, we
simply have found out that a zeroth order weak coupling approach is
insufficient to describe inflation. In the strong coupling regime
the dilaton may turn out to be non-constant after all.

It is interesting to note that our result confirms a conjecture in
\cite{Dine85b}, that a cosmological solution in which the
world is slowly sliding to its free Minkowski vacuum cannot be studied from this final state. From the reasoning in \cite{Dine85a,Dine85b} this
 appears to be a perfectly fine solution, if unlikely. Our result confirms their
expectation that such a slow-roll inflationary scenario is not
possible within tree-level worldsheet string theory.


To conclude: we have provided a proof of principle that the coarse
characteristics of the internal CFT determine whether and how
inflation occurs by expressing the slow-roll parameters in terms
of the beta functions of the internal CFT. As de Sitter-like
solutions only arise at one-loop in a Minkowski string worldsheet,
a necessary requirement for real and realistic worldsheet models
of string inflation is to include higher order string loop
corrections to the analysis.
This remains subject to further
investigation.

\paragraph{Acknowledgements}
We are grateful to A. Achucarro, J.O. Gong, S. Hardeman, M. Jackson, S. Sethi for numerous discussions and in particular to G. Shiu who was part of the initial collaboration. This research was supported
in part by a VIDI and a VICI Innovative Research Incentive Grant from the Netherlands Organisation for Scientific Research (NWO) and a Project Grant from the Dutch Foundation for Fundamental Research on Matter
(FOM).

\appendix
\section{Calculating the beta functions}\label{sec:betacalc}
In this appendix we will review the calculation of the beta functions (\ref{eq:betaeom}) of the total theory (\ref{eq:fullaction}). For more details concerning this calculation we refer to \cite{Freedman0510126}.

\subsection{Conformal perturbation theory}
For a general CFT $\mathcal{C}$ that is perturbed by adding operators
to the action,
\begin{equation}
S=S_{\mathcal{C}}+\int\d^2z\,u^I\mathcal{O}_I,
\end{equation}
the beta functions $\beta^I$ for the couplings $u^I$ can be defined as the coefficients of the trace of the stress-energy
tensor
\begin{equation}\label{eq:betafunctions}
\Theta=-\pi\beta^I\mathcal{O}_I.
\end{equation}
In the Zamolodchikov renormalization group scheme these can be
computed in an expansion in $u^I$ with $\delta_I=|\Delta_I-2|$
small
\cite{Kutasov88,Freedman0510126,Polchinski98},
\begin{equation}\label{eq:betapert}
\beta^I=(\Delta_I-2)u^I+2\pi C^{I}_{JK}u^Ju^K+\mathscr{O}(u^3),
\end{equation}
where $C^I_{JK}$ are the OPE coefficients defined
via
\begin{equation}\label{eq:OPEcoeff}
\mathcal{O}_J(z)\mathcal{O}_K(y)=\sum_IC^I_{JK}|z-y|^{\Delta_I-\Delta_J-\Delta_K}\mathcal{O}_I\left(\frac{z+y}{2}\right).
\end{equation}
In the coupled system $CFT_X\otimes CFT_{\mathcal{O}}$ that is deformed by the term
$S_{u(X)}=\int u^a(X)\mathcal{O}_a$ as described in the main text (\ref{eq:fullaction}), the operators in (\ref{eq:betafunctions}) are the three (types of) operators,
\begin{equation}
\mathcal{O}_G^{\mu\nu}=\frac{1}{2\pi\alpha'}\partial X^\mu\bar{\partial}X^\nu,\qquad
\mathcal{O}_a,\qquad
\mathcal{O}_\Phi=\frac{1}{8\pi}R^{(2)},
\end{equation}
which couple to the coupling functionals $G_{\mu\nu}(X)$, $u^a(X)$ and $\Phi(X)$ respectively. By a Fourier transform these coupling functionals may be seen as an infinite set of coupling constants $G_{\mu\nu}(p)$, $u^a(k)$ and $\Phi(q)$ that couple to the dressed operators $\mathcal{O}^{\mu\nu}_p=\frac{1}{2\pi\alpha'}\partial X^\mu\bar{\partial}X^\nu e^{ip\cdot X}\boldsymbol{1}$, $\mathcal{O}_{(k,a)}=\mathcal{O}_a e^{ik\cdot X}$ and $\mathcal{O}^\Phi_q=\frac{1}{8\pi}R^{(2)}e^{iq\cdot X}$ with dimensions
\begin{equation}\label{eq:scalingdimensions}
\Delta^G_{p}=2+\frac{\alpha'}{2}p^2,\qquad
\Delta_{(k,a)}=\Delta_a+\frac{\alpha'}{2}k^2,\qquad
\Delta^\Phi_q=2+\frac{\alpha'}{2}q^2.
\end{equation}
We are not constraining the graviton momentum or dilaton momentum to be lightlike. $p^2=0$ and $q^2=0$ would be the on-shell condition for a \emph{free} graviton and \emph{free} dilaton, whereas we wish to consider the coupled gravity-matter system. The OPE coefficients can be readily computed to be
\begin{subequations}\label{eq:interestingOPEcoeff}
\begin{eqnarray}
C^{(p,\boldsymbol{1})}_{(k_1,a)(k_2,b)}&=&-\frac{\alpha'}{8\pi}(k_1-k_2)_\mu(k_1-k_2)_\nu\delta^4\left(p-k_1-k_2\right)M_{ab},\\
C^{(k_1,a)}_{(k_2,b)(k_3,c)}&=&\delta^4\left(k_1-k_2-k_3\right)C^{a}_{bc},
\end{eqnarray}
\end{subequations}
where $C^a_{bc}$ are the OPE coefficients of the internal CFT and
we have denoted the Zamolodchikov metric by
$M_{ab}=4\pi^2C^{\boldsymbol{1}}_{ab}$.
Applying (\ref{eq:scalingdimensions}) and (\ref{eq:interestingOPEcoeff}) to
(\ref{eq:betapert}) and Fourier-transforming back to
position-space, yields
\begin{subequations}\label{eq:conformalperturbationbeta}
\begin{eqnarray}
\frac{1}{\alpha'}\beta^G_{\mu\nu}&=&-\frac{1}{2}\partial^\rho\partial_\rho G_{\mu\nu}+\frac{1}{2}M_{ab}\left(u^a\partial_\mu\partial_\nu u^b-\partial_\mu u^a\partial_\nu u^b\right),\\
\frac{1}{\alpha'}\beta^a&=&\frac{1}{\alpha'}\left((\Delta_a-2)u^a+2\pi C^a_{bc}u^bu^c\right)-\frac{1}{2}\partial^\rho\partial_\rho u^a\notag\\
&=&\frac{1}{\alpha'}\bar{\beta}^a(u)-\frac{1}{2}\partial^\rho\partial_\rho
u^a,\\
\frac{1}{\alpha'}\beta^\Phi&=&-\frac{1}{2}\partial^\rho\partial_\rho \Phi,\label{eq:conformalperturbationbetaPhi}
\end{eqnarray}
\end{subequations}
where we use (\ref{eq:betapert}) in reverse to express $\beta^a$ in terms of $\bar{\beta}^a$.

\subsection{Weyl anomaly and classical dilatonic contribution}
In addition to the operator effects from
(\ref{eq:conformalperturbationbetaPhi}), the beta function for
$\Phi$ receives a further contribution from the well-known Weyl
anomaly, a worldsheet contribution proportional to the worldsheet
curvature. Its contribution is determined by the central charge of
the spacetime non-linear sigma model as well as by that from the
(perturbed) internal theory as explained in the main text,
\begin{eqnarray}
\Theta_{1-loop}&=&-\left(\frac{c_X}{12}+\frac{1}{8}\bar{\beta}^\Phi\right)R^{(2)}
=-\pi\left(\frac{2c_X}{3}+\bar{\beta}^\Phi(u)\right)\frac{1}{8\pi}R^{(2)}\notag\\
&=&-\pi\left(\frac{2c_X}{3}+\bar{\beta}^\Phi(u)\right)\mathcal{O}_\Phi.
\end{eqnarray}
Comparing this expression with the definition of the beta
functions as coefficients in the stress-energy tensor
(\ref{eq:betafunctions}), we find a contribution
$\beta^\Phi_{1-loop}=\frac{2c_X}{3}+\bar{\beta}^\Phi(u)$ to the
beta function of the dilaton.

The final contribution to all of the beta functions comes from the
dilaton term (\ref{eq:dilaton}) in the worldsheet action, which
breaks Weyl invariance already at the classical level. Due to an
additional overall $\alpha'$-factor compared to the other terms in
the worldsheet, this classical contribution to the beta functions
is of the same order as loop effects from the classically Weyl
invariant terms. On a curved worldsheet the easiest way to
determine deviation from Weyl invariance is by calculating the
trace of the stress-energy tensor via
\begin{equation}
\Theta=\frac{-\pi}{\sqrt{g}}\frac{\delta S}{\delta g^{\alpha\beta}}g^{\alpha\beta}.
\end{equation}
This definition for $\Theta$ in terms of a variation of the
worldsheet metric differs by a factor from more common
definitions, which is necessary to relate the result properly with
our earlier definition (\ref{eq:betafunctions}). One can check
that this leads to the right result by looking at the metric and
dilaton field, whose contributions are well-known
\cite{Callan85,Polchinski98}. Making use of the equations of
motion for $X^\mu$,
\begin{equation}
\partial\bar{\partial}X^\rho=-\Gamma^{\rho}_{\mu\nu}\partial X^\mu\bar{\partial}X^\nu+\pi\alpha'\partial^\rho u^a\mathcal{O}_a+\frac{\alpha'}{8}\partial^\rho \Phi R^{(2)},
\end{equation}
the classical violation of Weyl invariance by the dilaton term (\ref{eq:dilaton}) is,
\begin{eqnarray}\label{eq:Thetaclassical}
\Theta_{classical}&=&\left.\frac{-\pi}{\sqrt{g}}\frac{\delta S_{\Phi(X)}}{\delta g^{\alpha\beta}}g^{\alpha\beta}\right|_{g_{z\bar{z}}=1/2}=-\partial\bar{\partial}\Phi(X)=-\left(\partial_\mu\partial_\nu\Phi\partial X^\mu\bar{\partial} X^\nu+\partial_\rho\Phi\partial\bar{\partial} X^\rho\right)\notag\\
&=&-\pi\left(2\alpha'\nabla_\mu\nabla_\nu\Phi\mathcal{O}^{\mu\nu}_G+\alpha'\nabla^\rho\Phi\nabla_\rho u^a\mathcal{O}_a+\alpha'(\nabla\Phi)^2\mathcal{O}_\Phi\right).
\end{eqnarray}
Again comparing with (\ref{eq:Thetaclassical}), we find contributions
\begin{subequations}
\begin{eqnarray}
\beta^G_{classical}&=&2\alpha'\nabla_\mu\nabla_\nu\Phi,\\
\beta^a_{classical}&=&\alpha'\partial^\rho\Phi\partial_\rho u^a,\\
\beta^\Phi_{classical}&=&\alpha'\partial^\rho\Phi\partial_\rho\Phi.
\end{eqnarray}
\end{subequations}

\subsection{Covariantization}
The beta functions $\beta^G_{\mu\nu}$, $\beta^a$, $\beta^\Phi$ found so far are (partially)
non-covariant. The terms obtained using conformal
perturbation methods are not spacetime covariant at first. This is inherent to the
conformal perturbation method, which uses correlation functions
defined with respect to \emph{flat} spacetime. Conformal perturbation is an expansion in $\delta G_{\mu\nu}=G_{\mu\nu}-\eta_{\mu\nu}$. If one corrects
for this by evaluating the Weyl transformation of all terms of the
coherent state of gravitons $G_{\mu\nu}(X)$, the expressions will
become covariant. Covariantization is necessary because the true
beta functions are gravitationally only consistent when all orders in $\delta G_{\mu\nu}$
are taking into account. Using background field methods one obtains the following spacetime covariant expressions \cite{Freedman0510126,Polchinski98},
\begin{subequations}
\begin{eqnarray}
\frac{1}{\alpha'}\beta^G_{\mu\nu}&=&R_{\mu\nu}+\frac{1}{2}M_{ab}\left(u^a\nabla_\mu\nabla_\nu u^b-\nabla_\mu u^a\nabla_\nu u^b\right)+2\nabla_\mu\nabla_\nu\Phi,\\
\frac{1}{\alpha'}\beta^a&=&\frac{1}{\alpha'}\bar{\beta}^a(u)-\frac{1}{2}\nabla^2 u^a+\nabla^\rho\Phi\nabla_\rho u^a,\\
\frac{1}{\alpha'}\beta^\Phi&=&U(u)-\frac{1}{2}\nabla^2\Phi+\left(\nabla\Phi\right)^2.
\end{eqnarray}
\end{subequations}

A further adaptation needs to be made to $\beta^a$, which is not covariant on the space of couplings $u^a(X)$. The right expression for $\beta^a$ should be
\begin{equation}\label{eq:covariantbetau}
\frac{1}{\alpha'}\beta^a=\frac{1}{\alpha'}\bar{\beta}^a(u)-\frac{1}{2}\nabla^2 u^a-\frac{1}{2}K^a_{bc}\nabla^\rho u^b\nabla_\rho u^c+\nabla^\rho\Phi\nabla_\rho u^a,
\end{equation}
where $K^{a}_{bc}$ is the connection coefficient associated
to the Zamolodchikov metric $M_{ab}$. In a general renormalization scheme it arises from contact terms
in the OPE. It has not appeared explicitly in the Zamolodchikov scheme because in that scheme $K^a_{bc}$ is already of first order in $u$ \cite{Freedman0510126}, as a result of which
$K^a_{bc}\nabla^\rho u^b(X)\nabla_\rho u^c(X)$ is beyond leading order in the calculation of the beta
functions. In the Zamolodchikov scheme (\ref{eq:covariantbetau}) is correct to leading order and by general covariance it holds in any
renormalization scheme.

The last step to obtain beta functions which are convenient to
work with, is to remove the double derivatives $\nabla\nabla u$ in
$\beta^G_{\mu\nu}$.\footnote{In \cite{Freedman0510126} this is
done by way of a diffeomorphism that is not entirely clear to the
authors.} This can be achieved by a shift in the dilaton
\begin{eqnarray}
  \label{eq:2}
  \Phi\rightarrow \Phi - \frac{1}{8} M_{ab}u^au^b
\end{eqnarray}
The term in $\beta^a$ and $\beta^\Phi$ that result from this shift are all of higher order, whereas  
 $\beta^G_{\mu\nu}$ precisely becomes
\begin{equation}
\frac{1}{\alpha'}\beta^G_{\mu\nu}=R_{\mu\nu}-M_{ab}\nabla_\mu u^a\nabla_\nu u^b+2\nabla_\mu\nabla_\nu\Phi+\frac{1}{2}e^{2\Phi}\nabla_{(\mu}\left(e^{-2\Phi}u^a\nabla_{\nu)} u^bM_{ab}\right),
\end{equation}
Here all covariantization on the scalar manifold of couplings with respect to $M_{ab}$ is implicit. 

\end{document}